\pdfoutput=1

\documentclass[aps,twocolumn,superscriptaddress,showkeys,prb]{revtex4-1}

\usepackage{amsmath}
\usepackage{graphicx}
\usepackage[bookmarks=false,colorlinks=true,linkcolor=blue,urlcolor=blue,citecolor=blue]{hyperref}

\makeatletter
\renewcommand{\fnum@figure}{Figure \thefigure}
\makeatother

\begin{document}

\title{Direct characterization of near-field coupling in gap plasmon-based metasurfaces}

\author{Rucha Deshpande}
\email{rad@mci.sdu.dk}
\affiliation{Centre for Nano Optics, University of Southern Denmark, Campusvej 55, DK-5230 Odense M, Denmark}

\author{Vladimir A. Zenin}
\author{Fei Ding}
\affiliation{Centre for Nano Optics, University of Southern Denmark, Campusvej 55, DK-5230 Odense M, Denmark}

\author{N. Asger Mortensen}
\author{Sergey I. Bozhevolnyi}
\affiliation{Centre for Nano Optics, University of Southern Denmark, Campusvej 55, DK-5230 Odense M, Denmark}
\affiliation{Danish Institute for Advanced Study, University of Southern Denmark, Campusvej 55, DK-5230 Odense M, Denmark}

\date{\today}


\begin{abstract}
Metasurfaces based on gap surface-plasmon resonators allow one to arbitrarily control the phase, amplitude and polarization of reflected light with high efficiency. However, the performance of densely-packed metasurfaces is reduced, often quite significantly, in comparison with simple analytical predictions. We argue that this reduction is mainly because of the near-field coupling between metasurface elements, which results in response from each element being different from the one anticipated by design simulations, which are commonly conducted for each individual element being placed in an artificial periodic arrangement. In order to study the influence of near-field coupling, we fabricate meta-elements of varying sizes arranged in quasi-periodic arrays so that the immediate environment of same size elements is different for those located in the middle and at the border of the arrays. We study the near-field using a phase-resolved scattering-type scanning near-field optical microscopy (s-SNOM) and conducting numerical simulations. By comparing the near-field maps from elements of the same size but different placements we evaluate the near-field coupling strength, which is found to be significant for large and densely packed elements. This technique is quite generic and can be used practically for any metasurface type in order to precisely measure the near-field response from each individual element and identify malfunctioning ones, providing feedback to their design and fabrication, thereby allowing one to improve the efficiency of the whole metasurface.
[This document is the unedited Author's version of a Submitted Work that was subsequently accepted for publication in \textit{Nano Letters}, \copyright American Chemical Society after peer review. To access the final edited and published work see \url{http://dx.doi.org/10.1021/acs.nanolett.8b02393}.]

\end{abstract}

\maketitle

\begin{figure}
\centering\includegraphics{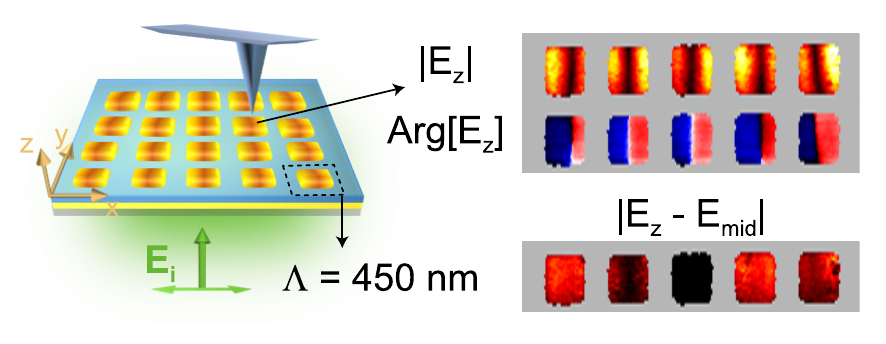}
\end{figure}

Metasurfaces are planar artificial nanostructures that can produce a desired optical response and realize a specific optical wavefront transformation by controlling multiple properties such as polarization, phase and amplitude for reflected and transmitted optical fields. Unlike bulk optical materials that control the propagation of light by gradual phase changes accumulated during the propagation through shaped and polished surfaces, metasurfaces can engineer the optical response by subwavelength periodic arrangement of meta-elements generating desired phase and amplitude profiles of scattered optical fields. The latter is achieved by gradually varying parameters of meta-elements (commonly placed in a subwavelength periodic arrangement) across a metasurface, introducing thereby local and different modifications in optical fields. Metasurfaces have become increasingly popular as they can be designed and fabricated to operate with relatively low losses, exhibiting numerous functionalities, for example, polarization splitting and detection\cite{SSun2012, AndersPors2013, APors2013-efficientphasecontrol, ZLi2015, APors2015-Stokes, APors2016-polarimetry, EMaguid2016, FDing2017, RDeshpande2017,FDing2017-review, FDingYiting2018, FDing2018-GSPmsreview}, waveplates\cite{APors2011, AAlu2011, ARoberts2012, APors2013qwplates, FDing-S2015}, lenses and focusing metamirrors\cite{SWang2017,SBoroviks2018}, random phase reflectors\cite{APors2016}, holograms\cite{WTChen2014, TZentgraph2015, PGenevet2015}, color printing\cite{ARoberts2014, XZhu2015, AKirstensen2016, HWang2017} and integrated multifunctional devices\cite{FDing2018, STang2018}.

The choice of meta-elements comprising the metasurface is the first step in realizing different applications. The meta-element represents a compact nanoantenna, which behaves as a near-resonant scatterer. By changing its parameters (shape, sizes and orientation) and periodicity, the resonances can be tuned to realize any phase change (within a full 2$\pi$ phase range) in the transmitted/reflected optical fields, which is a key prerequisite for the majority of applications. Most commonly, the optical response of an individual meta-element is predicted by a simplified approach, in which each element is considered to be placed in an artificial subwavelength periodic arrangement with identical neighboring elements. The rationale behind this simplification is that slowly varying gradient effects will not significantly affect the retrieved phase and amplitude response of the elements for reflected/transmitted light of desired wavelength and polarization. Then, based on the specific metasurface application, the elements are uniformly distributed with the same center-to-center separation as in the initial simulations of periodic arrays but, with the size of each nanoantenna comprising the element chosen to provide a desired phase profile. Thus, each element in the actual metasurface arrangement is no longer embedded in a periodic environment, which might render the actual response of meta-elements different from initial simulations, resulting in reduced efficiency of the whole metasurface. In order to enhance the efficiency, one may perform full simulations of a whole metasurface array, with all differently placed and sized elements being analyzed, and accurately calculate the expected phase and amplitude profile so as to finally optimize the metasurface design. However, such tremendous calculations would require powerful supercomputers, since the number of meta-elements can be very large depending on the application and yet, it does not directly verify if the meta-elements upon fabrication behave exactly as they are designed. It is important to note that the periodicity assumption remains the most basic assumption for the majority of metasurface designs, which requires optimization because the separations between the elements are within the range of dominant near-field coupling (usually of the order of $\lambda$/2$\pi$, where $\lambda$ is wavelength of operation)\cite{DWPohl1993}. Manifestation of the near-field coupling as the shift in resonance for varying period of arrays \cite{Hillenbrandnfcoupling2011, APors2013qwplates}, or for varying separation of dimer antennas\cite{CDeVault2017, TAtay2004,MStockmandimers2004,DSKimdimers2009,PabloAlonso-Gonzdimers2013} has been studied for different applications. However, for metasurfaces, the decrease in performance due to the near-field coupling has been reported by very few studies.\cite{CapassoNF2015, Yifat2014}.

In this work, we investigate near-field responses of each element of quasi-periodic arrays with respect to different environment and observe the influence of near-field coupling. We quantify the near-field coupling directly by using experimental near-field characterization techniques and support our findings further using simulations. For metasurface elements, we consider gap surface-plasmon (GSP) resonators, which can efficiently manipulate linearly polarized light.\cite{AndersPors2013, APors2013-efficientphasecontrol} The near-field characterization is performed using phase-resolved scattering-type scanning near-field optical microscopy (s-SNOM) in transmission mode. For simplicity, the study is performed for a single wavelength (1500 nm) and fixed linear polarization. We fabricate a quasi-periodic array of identical meta-elements in 5 columns, where the adjacent row elements have gradually varying widths. Thus, the elements from the middle column are effectively exposed to an infinite uniform periodic environment, compared to its neighbors, which is directly characterized using s-SNOM.    

To our knowledge, this is the first near-field characterization of GSP nanoantennas. By comparing the near-field maps from elements of the same size but different placement, we evaluate the near-field coupling, which is found to increase when increasing the element dimensions. We argue that the near-field coupling is proportional to the element size and inversely proportional to the center-to-center distance between elements (both size and distance are considered in the direction of the coupling). This conclusion is verified by additional experiments with increased separation, where no significant coupling was observed. Numerical calculations provide further insight into the dependence of this coupling on the wavelength and the distances between elements. Overall, the developed approach of near-field characterization is rather generic and can be applied for practically any metasurface type. It allows one to identify the elements with distorted phase and amplitude response due to the strong near-field coupling, so that each element parameters can individually be optimized to increase the efficiency of the whole metasurface. Moreover, our characterization technique can also be used for in-situ characterization of meta-elements in order to find and correct possible fabrication defects.

\section{Results and discussion}

\begin{figure}
	\centering
	\includegraphics{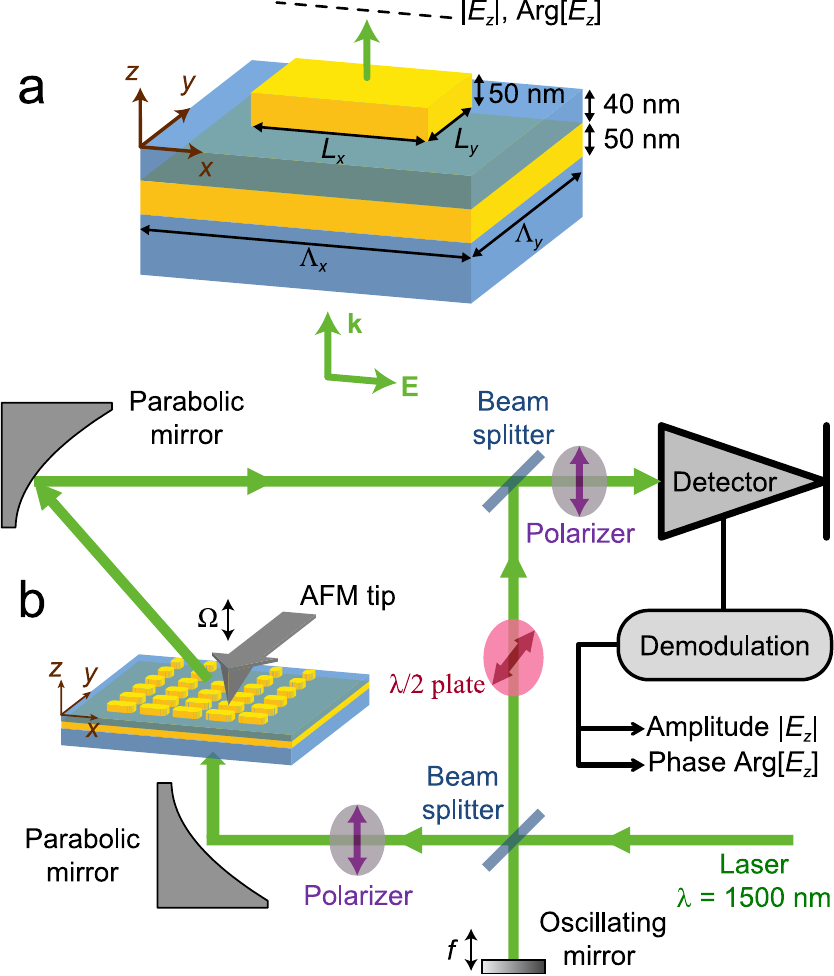}
	\caption{Configuration of gap plasmon elements and s-SNOM experimental setup. (a) Schematic of a gap plasmon-based configuration upon bottom normal incidence. (b) s-SNOM experimental setup, where the sample is illuminated from below with a defocused laser beam ($\sim$12 $\mu$m), polarized parallel to the nanobrick length $L_x$ (wavelength 1500 nm). The standard AFM metal-covered silicon tip scatters the near-field (predominantly its vertical component), and the scattered radiation, collected by the top parabolic mirror, is then mixed with the reference beam and interferometrically detected, yielding the near-field amplitude $\left|E_z\right|$ and phase ${\rm Arg}\left[E_z\right]$. Though transmittance through the 50 nm-thick gold film is only 1\%, the interference with a strong reference beam allowed a reliable detection of the near-field.}
	\label{fig1}
\end{figure}

\begin{figure*}
	\centering
	\includegraphics{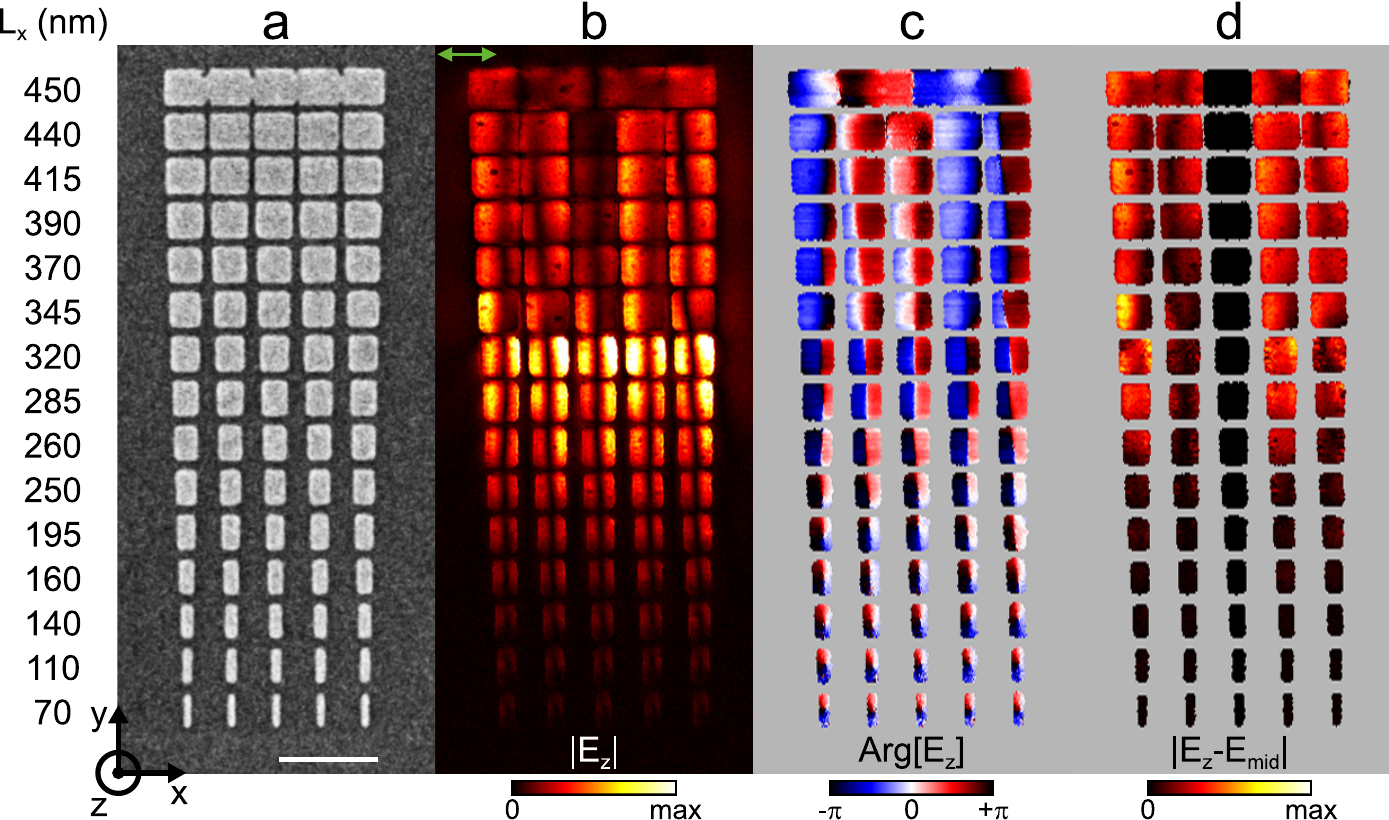}
	\caption{Experimental near-field characterization of quasi-periodic GSP array. (a) SEM image of fabricated meta-elements, whose measured length $L_x$ was fixed within each row and is indicated on the left, while their widths $L_y$ = 400 nm was kept constant for the whole array. Scale bar is 1 $\mu$m.  (b,c) Near-field (b) amplitude $\left|E_z\right|$ and (c) phase ${\rm Arg}\left[E_z\right]$, measured with s-SNOM. The incident light polarization is shown with green arrow in (b). (d) Amplitude of the difference $\left|E_z-E_{\rm mid}\right|$, where for each row the near-field of the middle-column element was subtracted from the near-field of elements from the other columns. Geometrical shape of elements, defined in the recorded topography, was used as a mask in (c) and (d).}
	\label{fig2}
\end{figure*}

Our GSP meta-elements are composed as following: a glass substrate, coated with 50 nm of gold, followed by a 40 nm-thin SiO\textsubscript{2} spacer layer and a top 50 nm-thick gold bricks (Figure~\ref{fig1}a). The planar gold bricks of length $L_x$ and width $L_y$ are arranged with a center-to-center separation of $\Lambda_x$ = $\Lambda_y$ = 450 nm. It is a commonly used configuration of GSP-based meta-elements except for the thickness of the bottom gold layer, which was chosen as a compromise: it should be thin enough to be partially transparent for further near-field measurements (transmittance $\sim$1 \%), while at the same time it should be thick enough to not significantly reduce the reflection ($\sim$97 \%).\cite{Bozhevolnyi2007} Numerical calculations of such elements in appropriate periodic boundary conditions and top light incidence (see Methods) have revealed that for the fixed nanobrick width $L_y$ and varied length $L_x$, the meta-element undergoes resonant behavior nearly in the whole 2$\pi$ phase range (see Methods and Supporting information, Figure~S1).

The near-field mapping was performed using the s-SNOM in the transmission mode, where the sample was illuminated normally from below at the wavelength of 1500 nm (Figure~\ref{fig1}b). The incident light was loosely focused ($\sim$12 $\mu$m diameter) and polarized along $x$-axis, which is along the nanobrick length $L_x$ (see Methods for more details). Though the illumination was from the bottom, the relative near-field response should be almost the same as that from the top normal illumination, since the GSP nanoantennas are optically thin. According to previous studies, the measured near-field most closely represents the normal near-field component $\left|E_z\right|$ approximately 50 nm above the sample surface \cite{Vzslotwaveguides2014, Vznanofocusing2015, Vzstripwaveguides2016}.

In order to observe the influence of the near-field coupling, we fabricated quasi-periodic arrays of GSP meta-elements with 5 columns, where the length of the elements $L_x$ is fixed within the row and it is gradually varied within the adjacent rows with a step of $\sim$25 nm, while the width of the elements $L_y$ is kept constant (Figure~\ref{fig2}a). In such a configuration, elements from the middle column will effectively experience close to periodic environment, while elements from the first and the last columns will experience the most uneven environment. The sample was fabricated using e-beam evaporation of gold, RF sputtering of SiO\textsubscript{2}, and a standard combination of the e-beam lithography with lift-off technique (see Methods). The recorded distributions of the near-field amplitude $\left|E_z\right|$ and phase ${\rm Arg}\left[E_z\right]$ for the quasi-periodic array with fixed width $L_y$ = 400 nm are presented in Figures~\ref{fig2}b,c. With the increase of $L_x$, the near-field response undergoes a resonance behavior, where the resonant length $L_x$ $\approx$ 320 nm can be identified by a maximum in the near-field amplitude (Figure~\ref{fig2}b). As for variations within each row, it is clearly seen by the naked eye that the near-field responses from meta-elements with small length $L_x$ are nearly identical for all 5 columns. However, starting from $L_x$ = 260 nm, the near-field response becomes different for elements belonging to different columns, and the difference can be observed both in the amplitude and phase distributions. In order to visualize this difference clearer, we subtract the near-field $\left|E_{\rm mid}\right|$ of the middle-column element from near-field maps of the elements from other columns (Figure~\ref{fig2}d). It is apparent then, that the near-field coupling is significant for meta-elements with $L_x > 285$ nm, because it causes such a strong difference in the near-field response from elements of different columns. The above conclusion is general and also holds for meta-elements with a different $L_y$ = 280 nm, and near-field maps for the corresponding quasi-periodic arrays can be found in the Supporting Information, Figure~S2.

\begin{figure}
	\centering
	\includegraphics{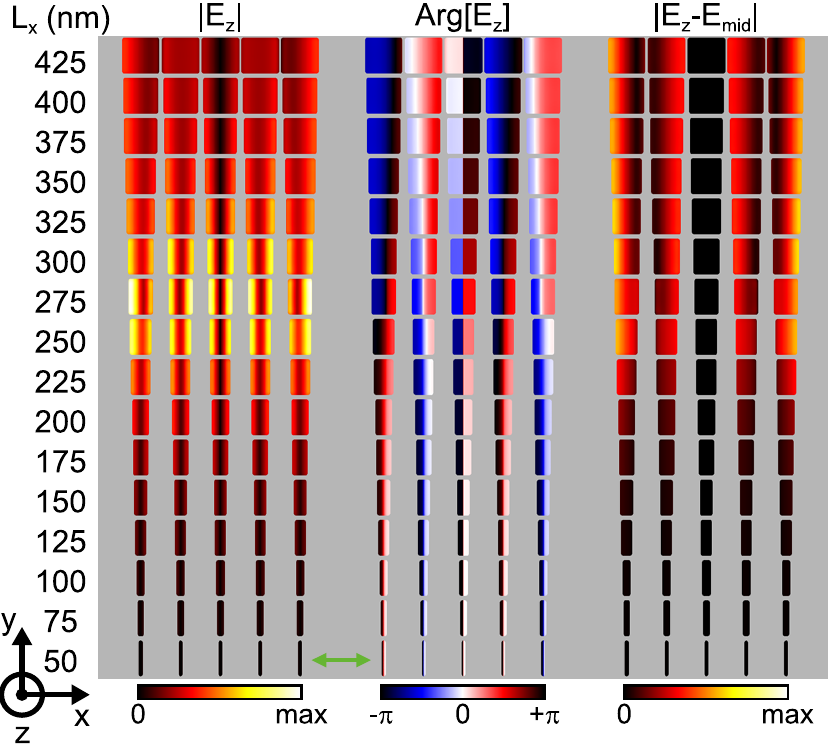}
	\caption{Simulated near-field results for quasi-periodic GSP array. The width of meta-elements $L_y$ is 400 nm, while their length $L_x$ is varied from 25 to 425 nm with 25 nm step, and it is shown on the left. The simulated near-field amplitude $\left|E_z\right|$, phase ${\rm Arg}\left[E_z\right]$, and the contrast with the middle element $\left|E_z-E_{\rm mid}\right|$ were calculated at the altitude of 50 nm above the top surface of meta-elements. The polarization of normally incident plane wave is shown with a green arrow. Designed geometrical cross-section of elements is used as a mask.}
	\label{fig3}
\end{figure}

To further strengthen our arguments, we performed numerical simulations in order to confirm, that the results are not affected by the inevitable inaccuracies associated with either fabrication or experimental limitations. A row of 5 identical GSP elements, distributed with a center-to-center distance of $\Lambda_x$ = $\Lambda_y$ = 450 nm, formed a simulation domain, which was truncated with periodic boundary conditions across $y$-axis and with perfectly matching layers in all other directions (see Methods section for more details). The near-field maps of such rows with different $L_x$ but the same $L_y$ = 400 nm, calculated at the altitude of 50 nm above the bricks, are shown in Figure~\ref{fig3}. The simulated results of the near-field amplitude $\left|E_z\right|$, phase ${\rm Arg}\left[E_z\right]$, and contrast with the middle element $\left|E_z-E_{\rm mid}\right|$ correspond well to the experimental results presented in Figure~\ref{fig2}. We notice that the elements with weak near-field coupling ($L_x < 250$ nm) have symmetric near-field distribution, featuring two lobes of the same amplitude but opposite phase. However, when the near-field amplitude becomes significant ($L_x > 250$ nm), the near-field distribution of non-middle elements is no longer symmetrical: the lobes of equal amplitude and opposite phase are no longer equal by the area, and the transition between phase lobes becomes gradual.

\begin{figure}
	\centering
	\includegraphics{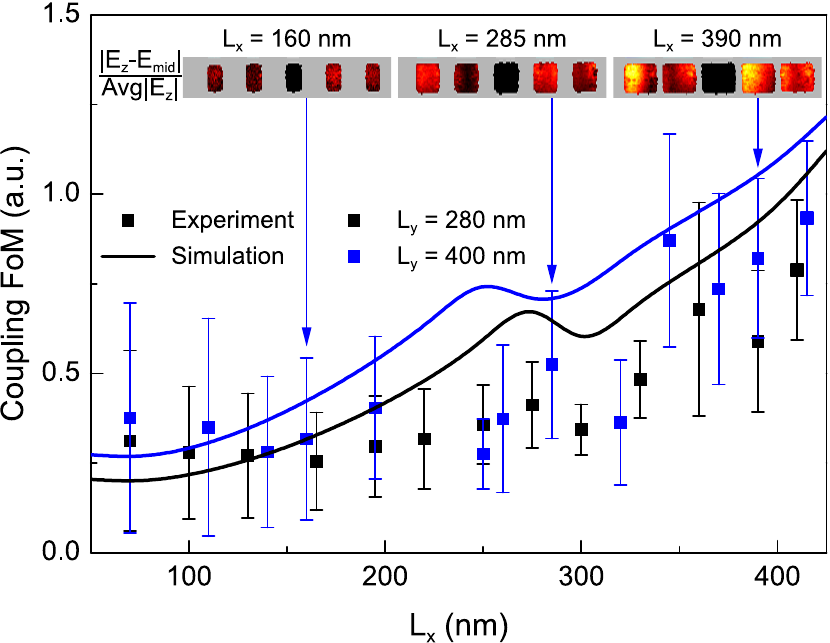}
	\caption{Near-field coupling figure of merit (FoM) for GSP meta-elements, arranged in the quasi-periodic arrays with a fixed $L_y$ of 280 (black) and 400 nm (blue). Simulated results are shown with lines, while experimental measurements are plotted with dots with error bars. Both the noise level of near-field maps and the variation of $\left|E_z-E_{\rm mid}\right|$ for different columns were taken into account for evaluation of the error. Coupling FoM within each row is calculated as ${\rm Avg}\left|E_z-E_{\rm mid}\right|/{\rm Avg}\left|E_z\right|$, where the first averaging is only within edge elements (from the first and the last columns), while second averaging is done for all 5 elements. Inset shows the normalized near-field contrast $\left|E_z-E_{\rm mid}\right|/{\rm Avg}\left|E_z\right|$ for case of three different lengths $L_x$ of 160, 285, 390 nm, demonstrating negligible, moderate, and strong near-field coupling, respectively.}
	\label{fig4}
\end{figure}

One may in the first place argue that low contrast $\left|E_z-E_{\rm mid}\right|$, observed in quasi-periodic arrays for small GSP element length ($L_x < 250$ nm) is due to its size being far from the resonant length ($L_x \approx$ 300 nm). Therefore, in order to take the resonance response into account, we first calculate the average near-field amplitude $\left|E_z\right|$ for five elements of the same row (i.e., the same $L_x$). Then, in order to evaluate the influence of the near-field coupling on the near-field response in a single Figure-of-Merit value (coupling FoM), we average the contrast $\left|E_z-E_{\rm mid}\right|$ over boundary elements from the first and the last columns and normalize it to the average $\left|E_z\right|$ for each row (Figure~\ref{fig4}). We observe an increase of the coupling FoM with an increase of the element length $L_x$. We notice a small distortion in otherwise gradual change of coupling FoM around $L_x \approx$ 300 nm, which can be attributed to the residual influence of the resonance behavior. The coupling FoM reaches a value of $\sim$0.5 at $L_x \approx$ 300 nm, after which it increases drastically. This feature indicates a transition beyond which the influence of the near-field coupling becomes significant.

\begin{figure*}
	\centering
	\includegraphics{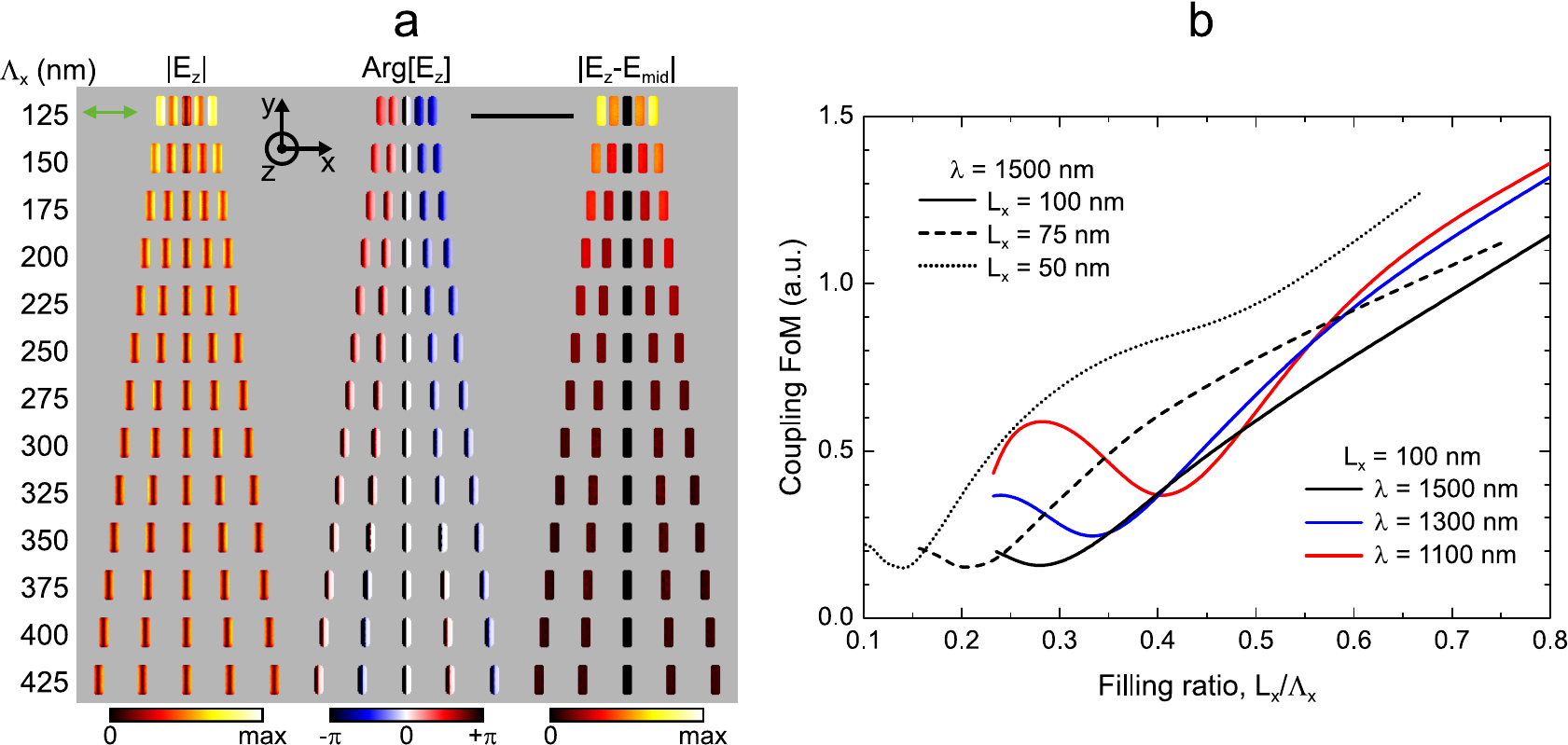}
	\caption{Numerical investigation of the near-field coupling in 5-column quasi-periodic GSP arrays with varied period along $x$-axis, $\Lambda_x$. (a) Simulated near-field amplitude $\left|E_z\right|$, phase ${\rm Arg}\left[E_z\right]$, and the contrast with the middle element $\left|E_z-E_{\rm mid}\right|$, calculated at the altitude of 50 nm above the top surface of meta-elements. Length $L_x$ and width $L_y$ was kept constant at 100 nm and 280 nm, correspondingly, while period along $x$-axis $\Lambda_x$ was varied from 125 to 425 nm (period along $y$-axis $\Lambda_y$ was fixed at 450 nm). The polarization of normally incident plane wave is shown with a green arrow. Scale bar: 1 $\mu$m. (b) Coupling FoM as a function of the filling ratio $L_x$/$\Lambda_x$. $L_x$ and $\lambda$ are indicated in the legend, while $L_y$ and $\Lambda_y$ were fixed at 280 and 450 nm, respectively.}
	\label{fig5}
\end{figure*}

In order to experimentally explore its possible relation to the GSP element length $L_x$, we fabricated and measured a further sample with a periodic GSP array, where identical element is repeated in 9 columns and 3 rows, changing the width gradually across every 3 rows. The periodicity was increased to $\Lambda_x$ = $\Lambda_y$ = 600 nm, which is ideal for $\lambda$ = 1500 nm since its resonant width is $\sim$300 nm. The results are shown in Supporting Information, Figure~S3. Compared to the quasi-periodic GSP arrays with $\Lambda$ = 450 nm, the identical GSP meta-elements in the periodic array with $\Lambda$ = 600 nm exhibit significantly less coupling. Thus, we suggest that the near-field coupling is proportional to the element length $L_x$ and inversely proportional to the center-to-center separation $\Lambda_x$. This can be explained by a simplified model with two spherical particles in the free space, where the influence of the near-field coupling $K$ is linearly proportional to the polarizability of each particle ($\alpha_1$ and $\alpha_2$) and inversely proportional to the sixth power of separation $R$ between them \cite{DWPohl1993}:
\begin{equation}
K \propto \frac{{\alpha}_1 {\alpha}_2}{{R}^6} \mkern\medmuskip \label{eqn1}
\end{equation}
In case of a GSP element its polarizability is proportional to the length $L_x$, however, in our study they are no longer in a free space. Nevertheless, we assume the same trends should be valid for the near-field coupling between GSP elements, and in the case of identical elements the coupling effect should be proportional to the length $L_x$ of each element and inversely proportional to their separation $\Lambda_x$:
\begin{equation}
K \propto \frac{{L_x}^a}{{\Lambda_x}^b} \mkern\medmuskip, \label{eqn2}
\end{equation}
where $a$ and $b$ are positive indices of power. In order to test the above hypothesis, we conducted additional simulations of GSP elements in 5-column quasi-periodic configuration, where separation $\Lambda_x$ was varied (Figure~\ref{fig5}). Near-field distributions for a set with fixed $L_x$ = 100 nm, $L_y$ = 280 nm, and $\Lambda_y$ = 450 nm are shown in Figure~\ref{fig5}a. One can clearly see how the near-field coupling influences both the amplitude and the phase response. The coupling FoM, calculated for similar sets with different $L_x$ and wavelength $\lambda$, are presented in Figure~\ref{fig5}b. The results are plotted as a function of the filling ratio $L_x$/$\Lambda_x$, which was chosen to test the assumption whether indices of power $a$ and $b$ in Equation~(\ref{eqn2}) are equal. However, numerical results in Figure~\ref{fig5}b demonstrate that the influence of the near-field coupling is not trivial (though a bit better agreement with Eq.~(\ref{eqn2}) was found assuming $a$ = 0.5 and $b$ = 1, see Supporting Information, Figure~S4).

\section{Conclusion}
In this work, we have investigated both experimentally and theoretically the near-field coupling between metasurface elements, which is usually not taken into account during the design stage and often resulting in detrimental effects in the metasurface performance. This problem occurs due to the fact that the design simulations are commonly conducted for each individual meta-element being placed in an artificial subwavelength periodic arrangement with identical neighboring elements, whereas elements in the actual metasurface are varying in size across the metasurface in order to generate a desired phase profile. Depending on the metasurface functionality, the size difference between neighbor elements can be very large, causing deviations (from the designed values) in the phase response of individual elements. This detrimental effect can be ascribed to the near-field coupling between adjacent meta-elements, that we have studied by fabricating and measuring quasi-periodic GSP arrays, in which the elements from the middle column are effectively exposed to a uniform periodic environment in stark contrast to the border elements having neighbors only from one side. 

The fabricated structures were experimentally characterized using the phase-resolved s-SNOM in the transmission mode, and theoretically considered using numerical simulations. By comparing the near-field maps from elements of the same size but different locations, the near-field coupling was evaluated and found to be proportional to the element length and inversely proportional to the center-to-center distance between elements (both size and distance are considered in the direction of the coupling).  Additional experiments and numerical simulations for different configurations verified this conclusion and provided further insight into the dependence of this coupling on the wavelength and distances between elements. Our technique of near-field characterization is particularly useful for inspection of individual elements in densely packed metasurfaces, which is impossible with any far-field methods. Considering the current trend in metasurfaces towards integrated functionalities, accurate response, and high efficiency, the findings of our study can be very useful in identifying the elements with distorted optical response, which can individually be further optimized to increase the efficiency of the whole metasurface. Additionally, our near-field characterization technique can also be used for inspection of meta-elements in order to find and correct possible fabrication defects.

\section*{Methods}

{\bf Fabrication.} We used Electron Beam Lithography (EBL) nanofabrication technique to fabricate the metasurface arrays. In this technique, thin layers of metals viz. gold, titanium are deposited using e-beam evaporation, while dielectric silicon dioxide (SiO\textsubscript{2}) spacer layer is deposited using RF sputtering. Adhesion is facilitated by deposition of 3 nm titanium within the layers. A positive resist 950 kDa poly(methyl methacrylate) (PMMA) is deposited onto the substrate coated with bottom gold layer (50 nm) and dielectric layer (40 nm) using spin coating to obtain a thickness of 100 nm. The resist is then exposed to the pattern designed for the nanobricks using scanning electron microscope (SEM, model: JEOL JSM-6490LV) with acceleration voltage of 30 kV, working distance of 9 mm, area dose of 200 $\mu$C/cm\textsuperscript{2}), write field of 30 $\times$ 30 $\mu$m, and step size of 2 nm. After exposure, the resist is developed for 30 s in a 3:1 mixture of isopropanol (IPA) and methyl isobutyl ketone (MIBK). The nanobricks of 50 nm height are fabricated by deposition of gold using e-beam evaporation and subsequent 10 hours incubation in PG-remover (commercially obtained solution) for lift-off of unexposed resist. The fabricated metasurface is imaged using scanning electron microscopy (SEM) in order to determine actual dimension of nanobricks (Figure~\ref{fig5}a). A non-uniform increase in the element length $L_x$ was caused by the proximity effect.

{\bf Numerical Simulations.} All modelings are performed using commercial finite element software Comsol Multiphysics (version 5.2), and a plane-wave excitation. Permittivity values of gold were taken from Johnson and Christy database, \cite{JohnsonandChristy1972} while the refractive index of glass substrate and SiO\textsubscript{2} spacer layer was assumed to be 1.45. The medium above the nanobricks is chosen to be air with refractive index of 1. In the simulations of individual GSP antenna (Supplementary Figure~S1) a single unit cell with periodic boundary conditions on the vertical sides of the cell was used. Excitation and collection ports were applied above and below the unit cell, followed by perfectly matched layers in order to minimize reflections. In the simulations of coupling (Figures~\ref{fig3}, \ref{fig5}, and S2) a unit cell with 5 GSP antennas was used, for which periodic boundary conditions were applied only on the $xz$ vertical sides of the cell because the experimentally investigated structure was quasi-periodic along $y$-axis with slowly varying $L_x$ length of GSP bricks. The remaining 4 boundaries of the unit cell were truncated with perfectly matched layers to minimize reflections. All edges of gold bricks were rounded with 10 nm radius of curvature. 

{\bf Near-field Microscopy.} We used commercial AFM-based scattering-type s-SNOM (Neaspec GmbH) with standard platinum-coated Si tips (Arrow\textsuperscript{TM} NCPt from NanoWorld) to measure near-field experimentally. The AFM tip was tapping with amplitude of $\sim$50 nm at frequency of $\sim$250 kHz. The sample was illuminated normally from below using a parabolic mirror to focus the light from a tunable telecom diode laser (TLB-6500-H-ES from New Focus, 1500 nm wavelength). The illumination spot size at the sample surface was estimated to be $\sim$12 $\mu$m in FWHM, thus, homogenously illuminating the structure over a large area. The illuminating beam travels in positive $z$-axis and its polarization is oriented along the varying nanobrick length $L_x$, i.e., $x$-axis. The scattered light was collected using a second parabolic mirror, placed above the tip. A resolution of both amplitude and phase was done by using a Mach-Zehnder interferometer with an oscillating mirror ($f \sim 300$ Hz) in the reference arm and a pseudo-heterodyne detection scheme.\cite{RHillenbrand2006} In order to remove background, the detected signal was demodulated at the third harmonic of the tip's tapping frequency. The results are presented in terms of s-SNOM measured near-field amplitude $\left|E_z\right|$ and phase ${\rm Arg}\left[E_z\right]$ maps. The unprocessed experimental data are shown in Figures~\ref{fig2}b,c and Supporting Information, Figures~S2-S3. 
 
\textbf{Acknowledgments}

The authors would like to thank for the support from European Research Council (Grant 341054, PLAQNAP), from VILLUM FONDEN (Grant 16498), and from University of Southern Denmark (SDU2020 funding).  

\bibliography{BiblNear-field}

\pagebreak
\widetext
\begin{center}
\textbf{\large Supporting information}
\end{center}
\setcounter{equation}{0}
\setcounter{figure}{0}
\setcounter{table}{0}
\setcounter{page}{1}
\makeatletter
\renewcommand{\thepage}{s\arabic{page}}
\renewcommand{\theequation}{S\arabic{equation}}
\renewcommand{\thefigure}{S\arabic{figure}}
\renewcommand{\bibnumfmt}[1]{[S#1]}
\renewcommand{\citenumfont}[1]{S#1}

\begin{figure}[hb]
	\centering
	\includegraphics{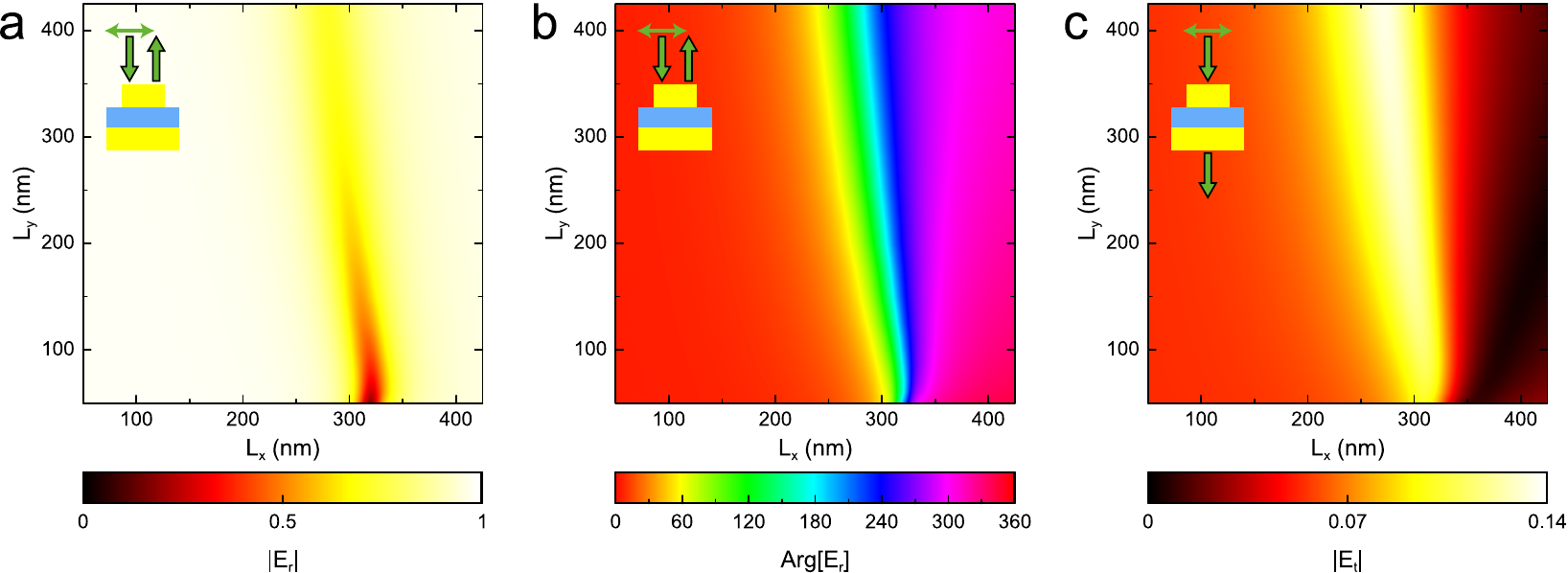}
	\caption{Simulated far-field response from GSP meta-element in periodic environment under normal top illumination. Simulated (a) amplitude $\left|E_{\rm r}\right|$ and (b) phase ${\rm Arg}\left[E_{\rm r}\right]$ of the reflection; and (c) amplitude of the transmission $\left|E_{\rm t}\right|$ upon normal illumination from the top.}
	\label{fig:sup-figure-1}
\end{figure}

\begin{figure}
	\centering
	\includegraphics{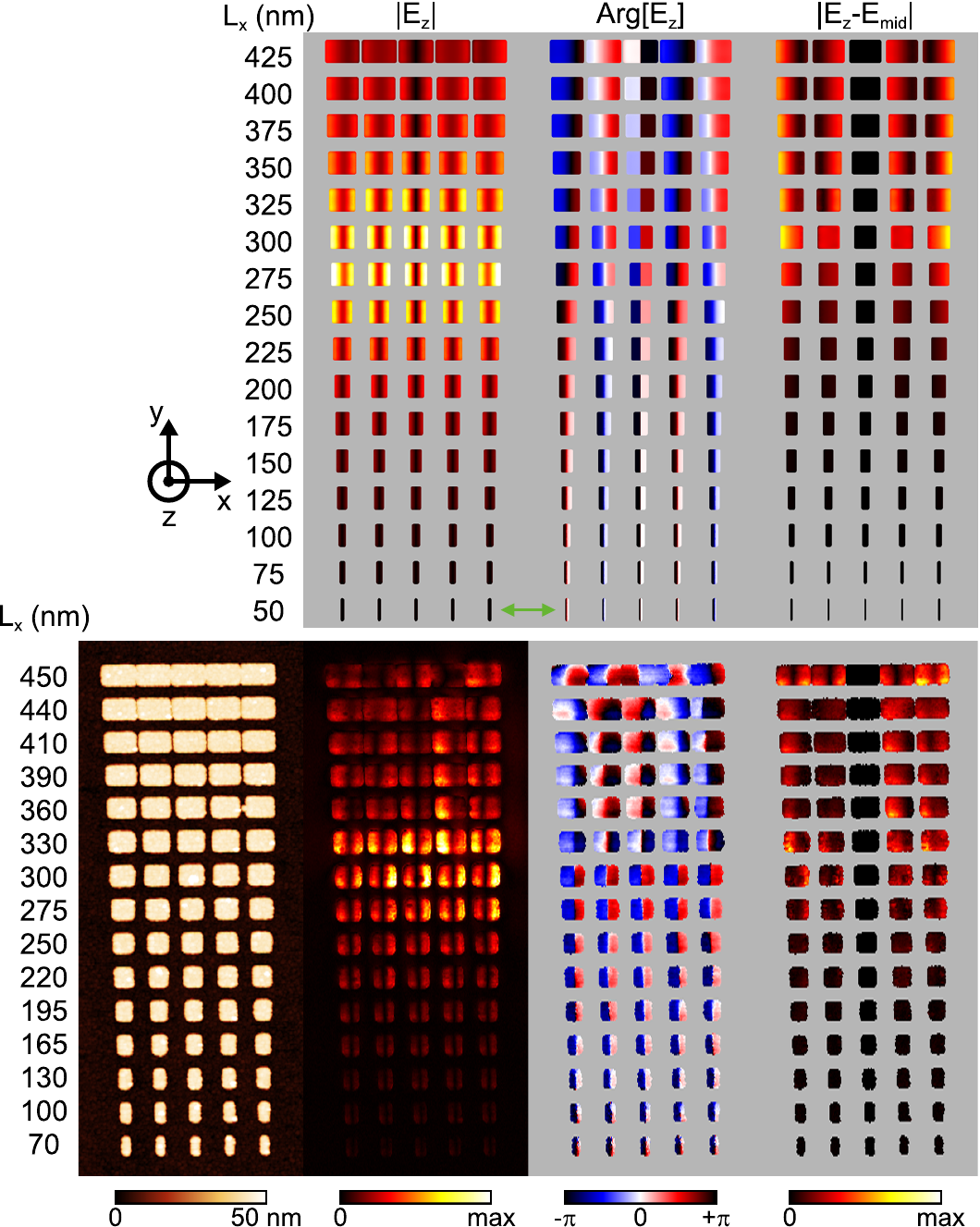}
	\caption{Near-field maps for quasi-periodic GSP array with $L_y$ = 280 nm. 
		Top: the simulated near-field amplitude $\left|E_z\right|$, phase ${\rm Arg}\left[E_z\right]$, and the contrast with the middle element $\left|E_z-E_{\rm mid}\right|$, calculated at the altitude of 50 nm above the top surface of meta-elements.
		Bottom: the experimentally measured s-SNOM topography, the near-field amplitude, phase, and the contrast, respectively. The polarization of normally incident plane wave is shown with a green arrow. The geometrical cross-section of elements is used as a mask.}
	\label{fig:sup-figure-2}
\end{figure}

\begin{figure}
	\centering
	\includegraphics{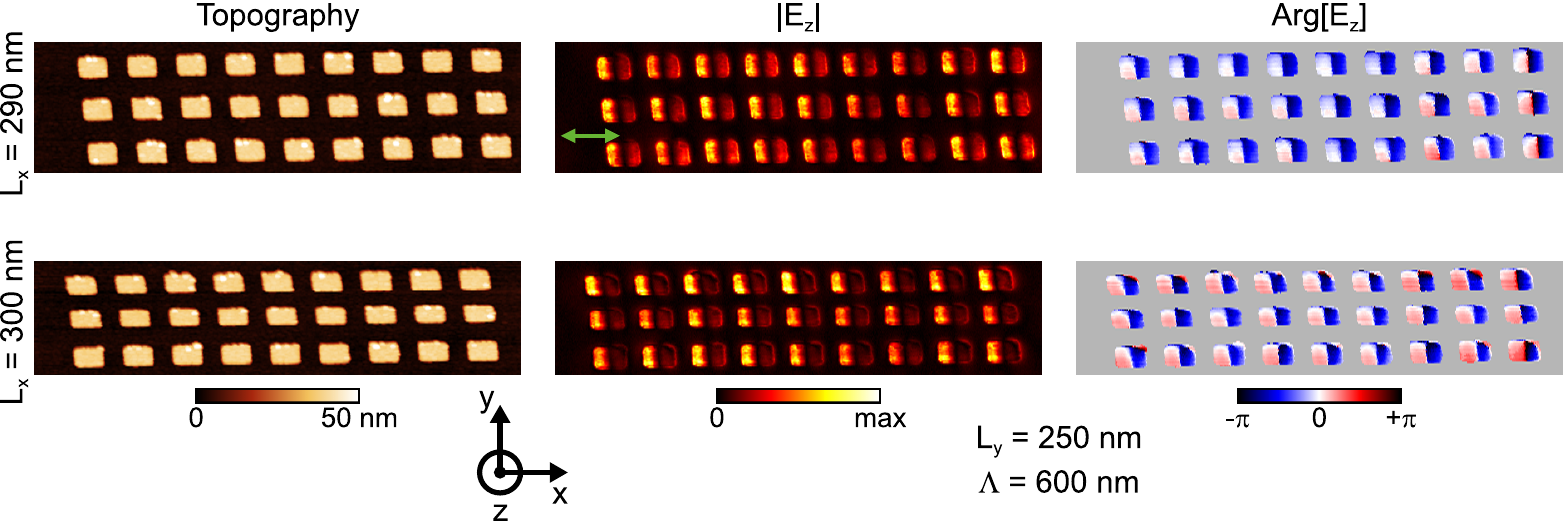}
	\caption{Experimental s-SNOM measured topography, near-field amplitude $\left|E_z\right|$ and phase ${\rm Arg}\left[E_z\right]$, measured for quasi-periodic array with identical elements being repeated into 9 columns and 3 rows (9$\times$3). The nanobrick length $L_x$ = 290 nm (top) and 300 nm (bottom), width $L_y$ = 250 nm, and periodicity $\Lambda$ = 600 nm ($\lambda$ = 1500 nm). The incident light polarization is shown with green arrow. Measurements were done with a different type of s-SNOM probe (uncoated Si tip), resulted in a different near-field distribution than is shown in Figures 2 and \ref{fig:sup-figure-2}. However, it is clear that the near-field coupling is decreased due to the increased periodicity.}
	\label{fig:sup-figure-3}
\end{figure}

\begin{figure}
	\centering
	\includegraphics{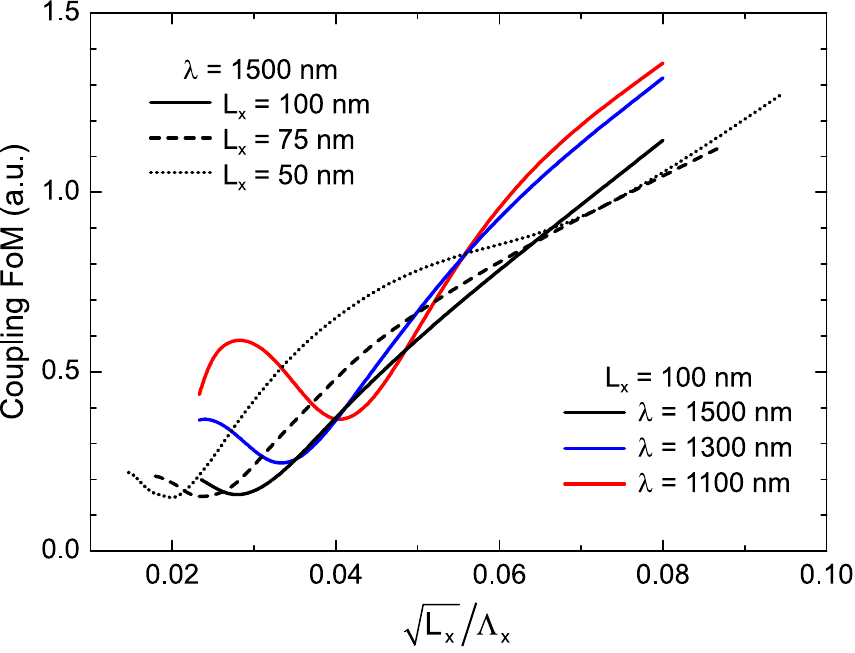}
	\caption{Numerical investigation of the near-field coupling in 5-column quasi-periodic GSP arrays with varied period along $x$-axis, $\Lambda_x$. The data is the same as in Figure 5 of the main text, but replotted as a function of the ratio $\sqrt{L_x}/\Lambda_x$. $L_x$ and $\lambda$ are indicated in the legend, while $L_y$ and $\Lambda_y$were fixed at 280 and 450 nm, respectively.}
	\label{fig:sup-figure-4}
\end{figure}

\end{document}